# Reconfigurable Logic Gates Using Single-Electron Spin Transistors

Pham Nam HAI[1], Satoshi SUGAHARA[2], and Masaaki TANAKA[1,3]

[1]*Department of Electronic Engineering, The University of Tokyo, 7-3-1 Hongo, Bunkyo-ku, Tokyo, 113-8656, Japan*

[2]*Imaging Science and Engineering Laboratory, Tokyo Institute of Technology, 4259-G2-14 Nagatsuta, Midori-ku, Yokohama, 226-8502, Japan*

[3]*SORST, Japan Science and Technology Agency, 4-1-8 Honcho, Kawaguchi, Saitama 332-0012, Japan*

We propose and numerically analyze novel reconfigurable logic gates using "single-electron spin transistors" (SESTs), which are single-electron transistors (SETs) with ferromagnetic electrodes and islands. The output characteristics of a SEST depend on the relative magnetization configuration of the ferromagnetic island with respect to the magnetization of the source and the drain, i.e., high current drive capability in parallel magnetization and low current drive capability in antiparallel magnetization. The summation of multiple input signals can be achieved by directly coupling multiple input gate electrodes to the SEST island, without using a floating gate. A Tucker-type inverter with a variable threshold voltage, a reconfigurable AND/OR logic gate, and a reconfigurable logic gate for all symmetric Boolean functions are proposed and simulated using the Monte Carlo method.

KEYWORDS: reconfigurable logic, spin transistor, single electron transistor, Coulomb blockade (CB), tunneling magnetoresistance (TMR)

## 1  Introduction

Using the spin degree of freedom in *active* semiconductor devices and integrated circuits is one of the most attractive new directions in spintronics research. Recently, we have proposed a new spin metal-oxide-semiconductor field-effect transistor (spin MOSFET), consisting of a MOS gate structure and half-metallic-ferromagnet contacts for the source and drain.[1] We also proposed novel spintronic reconfigurable logic gates employing spin MOSFETs.[2] We showed that all symmetric Boolean gates (AND, OR, XOR, NAND, NOR, XNOR, all-"0", and all-"1") could be realized using only five complementary MOS (CMOS) inverters including four spin MOSFETs. The functions of these circuits are nonvolatile and can be changed by changing the magnetization configurations of spin MOSFETs. This example shows the possibility of using the spin degree of freedom in *active* devices.

In this study, we propose a new kind of reconfigurable logic design using "single electron spin transistors" (SESTs), by which we can realize the analog input without using a floating gate. Figure 1 shows the schematic device structure and circuit notation of a SEST. The structure is similar to that of a single electron transistor (SET), which exploits the Coulomb blockade (CB) effect to switch the tunneling electric current.[3] The difference is that we use a ferromagnetic material for the island and electrodes for the SEST. The magnetization directions of the electrodes are fixed and maintained parallel to each other, whereas the magnetization of the island is reversible. Using the tunneling magnetoresistance (TMR) effect between the electrodes and the island, we can control the output current of the SEST by changing the magnetization direction of the island. When the magnetization of the island is parallel to that of the drain and the source electrode (parallel configuration; P), the tunneling resistances $R_{1-P}$ and $R_{2-P}$ are small; thus, the output current (drain current) is large. On the other hand, when the magnetization of the island is antiparallel to that of the drain and the source electrode (antiparallel configuration; AP), the tunneling resistances $R_{1-AP}$ and $R_{2-AP}$ are large; thus, the output current is small. Except for the TMR effect, the operating principle of the SEST is basically the same as that of the SET. The summation of the multiple input signals can be achieved by directly coupling the multiple input gate electrodes to the SEST islands, without using a floating gate. This is possible because the electrical characteristics of the SEST are determined by the total number of electrons on the counter electrodes of the island. This advantage, as well as the initial ultrasmall size of the SEST, makes circuits easy to be scaled down. Another advantage of the SEST is its high magneto-current ratio, which originates from the tunneling nature of the drain-source current.

In the following sections, we show the operation principle of a SEST, and then we present reconfigurable logic designs using SESTs; a Tucker-type inverter with a variable threshold voltage, a reconfigurable AND/OR logic gate, and a reconfigurable logic gate for two-input all symmetric Boolean functions. We show that these circuits can be designed using just four SESTs and six SETs without any floating gates. We theoretically analyze and show the operation of these reconfigurable logic gates by Monte Carlo simulations.[4]

## 2  Operation of Single SEST

In this section, we show the calculated operating characteristics of a SEST. The parameters used in the calculations are $C_1 = 9.9$ aF, $C_2 = 0.1$ aF, $C_g = 0.1$ aF, $R_{1-P} = 9.9$ MΩ, $R_{1-AP} = 60.8$ MΩ, $R_{2-P} = 0.1$ MΩ, $R_{2-AP} = 0.95$ MΩ, and $T = 0.01 e^2 /(C_1 + C_2 + C_g)k_B = 0.92$ K. Here, $C_1$, $C_2$, and $C_g$ are the capacitances of the first tunnel junction, second tunnel junction, and gate electrode with respect to the island, respectively, as shown in Fig. 1; $T$ is the temperature and $k_B$ is the Boltzmann's constant.

Details of the calculation are shown in Appendix A.

Figure 2 shows $I_{ds}$-$V_{ds}$ and $I_{ds}$-$V_{gs}$ characteristics at different magnetization configurations of a SEST, where $I_{ds}$, $V_{ds}$, and $V_{gs}$ are the drain-source current, drain-source voltage, and gate-source voltage, respectively. The $I_{ds}$-$V_{ds}$ characteristics show Coulomb gaps and the $I_{ds}$-$V_{gs}$ characteristics show Coulomb oscillations, similarly to those of a SET. Steplike features called Coulomb staircases also appear clearly in Fig. 2(a) due to the strong asymmetry of the chosen parameters between the source-island and the island-drain tunnel junctions.[5] Both the $I_{ds}$-$V_{ds}$ and $I_{ds}$-$V_{gs}$ characteristics have large output currents in P magnetization and low output currents in AP magnetization, as expected. The oscillating currents in Fig. 2(b) can be shifted along the $V_{gs}$ axis by applying a voltage on the control gate coupled to the island through the control capacitor.[5] This ability allows us to design the function of SEST as an n-type transistor or a p-type transistor by simply adjusting the voltage of the control gate.

We emphasize that the TMR effect is not the only method to change the resistance of the tunnel junctions by changing the magnetization direction of the electrodes. It is also possible to change the resistance of tunnel junctions using the anisotropy magnetoresistance (AMR) effect, which also causes a huge change in tunneling resistances under the CB regime.[6] The SEST may include all kinds of single electron transistor whose tunneling resistances depend strongly on the magnetization directions of the ferromagnetic electrodes. In the following sections, we assume that the tunneling resistances are changed by the TMR effect.

## 3    Tucker-Type Inverter with Variable Threshold Voltage

Figure 3 shows a Tucker-type inverter with a variable threshold voltage. The inverter is composed of two SESTs (SEST 1 and SEST 2). The structures of these two SESTs are exactly the same. The control gate of SEST 1 is connected to the ground so that it acts as a p-type transistor, i.e., when the gate voltage $V_{in}$ is "1", the transistor SEST 1 is OFF and vice-versa. On the other hand, the control gate of SEST 2 is connected to the supply voltage so that it acts as an n-type transistor, i.e., when the gate voltage $V_{in}$ is "1", the transistor SEST 2 is ON and vice-versa.[7] In Fig. 4, we show the normalized transfer characteristics ($V_{out}/V_{dd}$ vs $V_{in}/V_{dd}$) of this Tucker-type inverter, whose threshold voltage can be changed by the magnetization configuration. The parameters used in the calculations are $C_1 = 1$ aF, $C_2 = 2$ aF, $C_g = 8$ aF, $C_b = 7$ aF, $C_{out} = 24$ aF, $R_{1-P} = R_{2-P} = 0.5$ MΩ, $R_{1-AP} = R_{2-AP} = 5$ MΩ, $V_{dd} = 7$ mV, and $T = 0.001 e^2 /(C_1 + C_2 + C_g + C_b)k_B = 52$ mK. Here, $C_b$ and $C_{out}$ are the capacitances of the control gates and output terminal, respectively. Details of the calculation are shown in Appendix A. The blue curve shows the transfer characteristic when SEST 1 is in P magnetization and SEST 2 is in AP magnetization ({SEST 1, SEST 2} = {P, AP}). The red curve shows the transfer characteristic when SEST 1 is in AP magnetization and SEST 2 is in P magnetization ({SEST 1, SEST 2} = {AP, P}). In the region where $V_{in}$ is smaller than the threshold voltage $V_{th-SEST}$ of SEST 2 or larger than the threshold voltage $V_{dd} - V_{th-SEST}$ of SEST 1 (the gray shaded area in Fig. 4), either SEST 2 or SEST 1 is OFF; thus, the output current does not flow. In this region, the output voltage does not depend on the magnetization configurations. However, the situation is different in the transient region (the white area in Fig. 4) where both SEST 2 and SEST 1 are ON. In the magnetization configuration of {SEST 1, SEST 2} = {P, AP}, the resistances of the tunnel junctions of SEST 1 are smaller than those of SEST 2; thus, the output voltage is pulled up near the supply voltage ($V_{dd}$). In the magnetization configuration of {SEST 1, SEST 2} = {AP, P}, the output voltage is pulled down near the ground level. In other words, we can change the threshold voltage of the inverter by changing the magnetization configuration of the circuit.

## 4   Reconfigurable AND/OR Gate

The first application of the Tucker-type inverter with a variable threshold voltage is a reconfigurable AND/OR gate, as shown in Fig. 5. The circuit is composed of two stages. The first stage, as shown in the yellow area of Fig. 5, is a two-input Tucker-type inverter with a variable threshold voltage and is composed of two SESTs. Two-input gate electrodes $V_A$ and $V_B$ are directly coupled to each island of SESTs through two capacitors whose capacitances are half those of the Tucker-type inverter described in §3. This design creates an effective analog input $V_{in}$ to both SESTs from two digital inputs $V_A$ and $V_B$ by the relation $V_{in} = (V_A + V_B) / 2$. Consequently, the effective analog input $V_{in}$ values corresponding to digital inputs "00", "01" or "10", and "11" of $V_A$ and $V_B$ are "0", "1/2", and "1", respectively. As described in §3, the output $V_m$ of the first stage is "1" and "0" when $V_{in}$ = "0" and "1", respectively, irrespective of the magnetization configurations of the two SESTs. However, when $V_{in}$ is "1/2", $V_m$ is weak "1" and weak "0" when the magnetization configurations of SESTs are {SEST 1, SEST 2} = {P, AP} and {SEST 1, SEST 2} = {AP, P}, respectively. The second stage of the circuit, as shown by the green area of Fig. 5, is a normal Tucker-type inverter composed of two SETs and works as an inverse amplifier for $V_m$. This second stage inverses weak "0" / weak "1" to strong "1" / strong "0" when the input $V_{in}$ is "1/2". Note that the second stage can be constructed by a normal CMOS-based inverter as well, but a SET-based inverter is preferable because its input capacitance is sufficiently small to cover the low current driving capability of SESTs. The truth table of the circuit for each magnetization configuration is shown in Table I. By changing the circuit magnetization configuration from {SEST 1, SEST 2} = {P, AP} to {SEST 1, SEST 2} = {AP, P}, we can reconfigure the function of the circuit from the AND gate to the OR gate.

Figure 6 shows an example of the AND/OR circuit with parameters and its static operation calculated by Monte Carlo simulation. The parameters used in the simulations are $C_1 = 1$ aF, $C_2 = 2$ aF, $C_g = 8$ aF, $C_b = 7$ aF, $C_{out} = 24$ aF, $V_{dd} = 7$ mV, $T = 0.001e^2/(C_1+C_2+C_g+C_b)k_B = 52$ mK, $R_{1\text{-AP}} = R_{2\text{-AP}} = 5$ MΩ, and $R_{1\text{-P}} = R_{2\text{-P}} = 0.5$ MΩ for SEST 1 and SEST 2. As shown in Figs. 6(b) and 6(c), the circuit works as designed.

This example shows that it is possible to construct reconfigurable logic gates using only SESTs and SETs without using any floating gate. Consequently, we are able to not only scale down the circuits but also keep the operating speed at an acceptable level. For the circuit in Fig. 5, the time constant of the output $V_m$ of the first stage is given by $R_{eff}(C_{out}+C_{ext})$, where $C_{ext}$ is the input capacitance of the second stage, and $R_{eff}$ is the effective resistance of SESTs,[7] which is given by

$$R_{eff} = \frac{C_1+C_2+C_g+C_b}{C_2} R_{1\text{-ap}} + \frac{C_1+C_2+C_g+C_b}{C_1+C_g+C_b} R_{2\text{-ap}}. \tag{1}$$

The time constant of $V_m$ of the circuit shown in Fig. 6(a) is about 1.7 ns.

Let us then discuss about the possibility of fabricating SEST that can work at room temperature using conventional materials. In the above calculation, an idealized TMR ratio = $(R_{AP}-R_P)/R_P$ of 900% for each tunnel junction is assumed. In order to estimate the lowest TMR ratio at which the circuit can work, we have calculated the output voltages of the above-mentioned circuit with different TMR ratios. We varied the TMR ratio by changing $R_{AP}$ with fixed $R_P$. Figures 7(a) and 7(b) show the output voltages of the AND and OR gate at different TMR ratios, respectively. The other parameters are the same as those in Fig. 6. The output voltages do not depend on the TMR ratio when $\{V_A, V_B\} = \{0, 0\}$ and $\{V_A, V_B\} = \{1, 1\}$. However, when $\{V_A, V_B\} = \{0, 1\}$ and $\{V_A, V_B\} = \{1, 0\}$, the output voltages increase for the AND gate, and decrease for the OR gate with decreasing TMR ratio. This is because when $\{V_A, V_B\} = \{0, 1\}$ and $\{V_A, V_B\} = \{1, 0\}$, the output voltages depend on the relative magnitudes of $R_\text{AP}$ and $R_\text{P}$. If the high (low) voltage margins are set as $V_\text{out} > 5$ mV ($V_\text{out} < 2$ mV), the smallest TMR ratio $TMR_\text{min}$ at which the circuit still works is about 150 %. $TMR_\text{min}$ can be reduced more using an inverse amplifier with a sharper high-to-low transition of the $V_\text{out}$-$V_\text{in}$ curve. TMR ratios larger than 150% have been achieved experimentally by engineering the wave functions of the tunneling electrons in magnetic tunnel junctions using conventional FM electrodes and a single crystal tunnel barrier.[8-11]

The fluctuation of the tunnel resistances may also set limitation on the operation of the circuit, since it results in an effective TMR ratio fluctuation from the expected value by two ways. Firstly, the TMR ratio may be changed due to the barrier thickness dependence of TMR. Secondly, the TMR ratio can be changed directly due to the changes in $R_\text{AP}$ and $R_\text{P}$. Because the tunnel resistances change exponentially with the tunnel barrier thickness, the latter is much larger than the former. Thus, in the following argument, we only discuss the TMR ratio fluctuation caused by the latter. Based on the calculation described in Appendix B, we found the minimum TMR ratio $TMR_\text{min}'$ needed to overcome the resistance fluctuation. $TMR_\text{min}'$ is given by

$$TMR_\text{min}' = \left(TMR_\text{min} + \left|\frac{\Delta R}{R}\right|\right)\left(1 - \left|\frac{\Delta R}{R}\right|\right). \qquad (2)$$

Here, $\Delta R/R$ is the resistance fluctuation. Consequently, when the fluctuation of the tunnel resistances exists, a larger minimum TMR ratio is required. For $|\Delta R/R|=10\%$, $TMR_\text{min}'$ is 178%. Fortunately, $|\Delta R/R|$ as small as 2% has been realized at the mass production level by accurately controlling the tunnel barrier thickness of the vertical magnetic tunnel junctions.[12] For such a small $|\Delta R/R|$, $TMR_\text{min}'$ is 155 %, which is not much larger than $TMR_\text{min}$.

The high-temperature operation of SEST can be obtained by reducing the sizes of the source, drain, and island to a few nanometers while keeping the thermal stability factor $K_\mathrm{u}V/k_\mathrm{B}T$ larger than about 50 for 10 year nonvolatility, where $K_\mathrm{u}$ is the magnetic anisotropy constant and $V$ is the volume of the ferromagnetic island. Such a condition can be achieved using magnetic shape anisotropy or materials with large magneto-crystalline anisotropy, such as FePt and CoPt with a $K_\mathrm{u}$ range of 5–7×10$^7$ erg/cm$^3$.[13] Magnetization switching of FM islands having a large coercive field can be achieved using the spin torque transfer technique.[14,15] Note that the technologies for fabricating small SETs working at room temperature have already been well established.[16-19] The integration of SETs and their room temperature operation has also been demonstrated.[20] Analog pattern matching using integrated SET circuit has been performed at room temperature.[21] Consequently, the fabrication of SESTs and their integrated circuits working at room temperature is not technologically unreachable.

5  Reconfigurable Logic Gate for All Symmetric Boolean Functions

In §4, the output is programmable only at $V_\mathrm{in}$ = "1/2"; thus, reconfigurable functions are limited to the AND gate and OR gates. In this section, we show an extended version of the AND/OR gate whose output can be programmable at every input. This gate can perform any two-input symmetric Boolean functions (AND, OR, XOR, NAND, NOR, XNOR, all-"1", and all-"0"), as shown below. Figure 8 shows the circuit configuration of this reconfigurable gate. The circuit is composed of two stages. The first stage is composed of four SESTs (SEST 1, SEST 2, SEST 3, and SEST 4) and two SET based inverters (INV 1 and INV 2). SEST 1 and SEST 3 work as p-type transistors, whereas SEST 2 and SEST 4 work as n-type transistors. The inverter INV 1 has a threshold voltage higher than "1/2" and its output electrode is connected to the gate electrode of SEST 3. The inverter INV 2 has a threshold voltage lower than "1/2" and its output electrode is connected to the gate electrode of SEST 4. Two digital inputs $V_\mathrm{A}$ and $V_B$ are coupled to each island of SEST 1, SEST 2, INV 1, and INV 2 through two capacitors whose capacitances are half those of the Tucker-type inverter described in § 3, so that the inputs $V_{in}$ to these four devices are $V_{in} = (V_A + V_B) / 2$. The second stage is a SET-based inverter (INV 3) with a threshold voltage of "1/2" working as an inverse amplifier for the output $V_m$ of the first stage. The operation of the circuit is as follows.

(1) When $\{V_A, V_B\} = \{1, 0\}$ or $\{0, 1\}$, the input $V_{in}$ is "1/2", thus SEST 1 and SEST 2 are ON, whereas SEST 3 and SEST 4 are OFF. In this case, the circuit works in the same way as the AND/OR gate in Fig. 5. When {SEST 1, SEST 2} = {P, AP}, the output is "0", and when {SEST1, SEST} = {AP, P}, the output is "1".

(2) When $\{V_A, V_B\} = \{0, 0\}$, the input $V_{in}$ is "0", thus SEST 1 and SEST 4 are ON, whereas SEST 2 and SEST 3 are OFF. In this case, the electric current flows from the supply voltage to the ground through SEST 1 and SEST 4. By setting $R_{\text{SEST 4-P}} < R_{\text{SEST 1-P}} < R_{\text{SEST 1-AP}} < R_{\text{SEST 1-AP}} < R_{\text{SEST 4-AP}}$, the output voltage $V_{out}$ is "1" and "0" when the magnetization configuration of SEST 4 is P and AP, respectively. Here $R_{\text{SEST i-P}}$ and $R_{\text{SEST i-AP}}$ are the resistance of the tunnel junctions of SEST $i$ in P and AP magnetization configurations, respectively.

(3) When $\{V_A, V_B\} = \{1,1\}$, the input $V_{in}$ is "1", thus SEST 2 and SEST 3 are ON, whereas SEST 1 and SEST 4 are OFF. In this case, the electric current flows from the supply voltage to the ground through SEST 2 and SEST 3. By setting $R_{\text{SEST 3-P}} < R_{\text{SEST 2-P}} < R_{\text{SEST 2-AP}} < R_{\text{SEST 3-AP}} < R_{\text{SEST 4-AP}}$, the output voltage $V_{out}$ is "1" and "0" when the magnetization configuration of SEST 3 is AP and P, respectively.

The truth table of this reconfigurable circuit for each magnetization configuration is shown in Table II. An example of the circuit with parameters and its operation calculated by Monte Carlo simulations are shown in Fig. 9. The parameters used in the simulations are $C_1 = 1$ aF, $C_2 = 2$ aF, $C_g = 8$ aF, $C_b = 7$ aF, $C_{out} = 24$ aF, $V_{dd} =$

7 mV, $T = 0.001e^2/(C_1 + C_2 + C_g + C_b)k_B$ = 52 mK, $R_{1\text{-AP}} = R_{2\text{-AP}}$ = 50 MΩ, and $R_{1\text{-P}} = R_{2\text{-P}}$ = 5 MΩ for SEST 1 and SEST 2, and $R_{3\text{-AP}} = R_{4\text{-AP}}$ = 500 MΩ and $R_{3\text{-P}} = R_{4\text{-P}}$ = 0.5 MΩ for SEST 3 and SEST 4. As shown in Fig. 9(b), all two-input symmetric Boolean functions are operated as designed.

## 6.  Conclusion

We propose a new kind of reconfigurable logic design using single electron spin transistors (SESTs), by which we can realize analog inputs without using a floating gate. A Tucker-type inverter, an AND/OR reconfigurable gate, and a reconfigurable logic gate for two-input all symmetric Boolean functions were proposed and their operations were calculated by Monte Carlo simulations. The proposed logic gates can provide nonvolatile and scalable reconfigurable hardware for future electronics.

**Acknowledgments**

This work was supported by the Special Coordination Programs for Promoting Science and Technology, Grant-in-Aids for Scientific Research from the Ministry of Education, Culture, Sports, Science and Technology, and Kurata Memorial Hitachi Science & Technology Foundation. One of the authors (P. N. Hai) acknowledges support from the Japan Society for the Promotion of Science for Young Scientists.

**Appendix A**

Here, we describe the calculation procedure for the $I_{ds}$-$V_{ds}$ and $I_{ds}$-$V_{gs}$ characteristics of a single SEST and for the static operation of reconfigurable logic gates using SESTs. In the calculation, we assumed that the tunneling rate through every junction is much smaller than the rate of spin relaxation, so that the chemical potential in every island is spin-independent. This leads to a master equation that can be used to describe the time evolution of the probability $p_n(t)$ for $n$ excess electrons in a particular island:

$$\frac{dp_n}{dt} = \Gamma_{n+1 \to n} p_{n+1} + \Gamma_{n-1 \to n} p_{n-1} - (\Gamma_{n \to n+1} + \Gamma_{n \to n-1}) p_n, \quad (A\cdot 1)$$

where the net tunnel rates $\Gamma_{n \to n+1}$ and $\Gamma_{n \to n-1}$ are given by

$$\Gamma_{n \to n+1} = \vec{\Gamma}_L(n) + \vec{\Gamma}_R(n), \quad (A\cdot 2)$$

$$\Gamma_{n \to n-1} = \overleftarrow{\Gamma}_L(n) + \overleftarrow{\Gamma}_R(n). \quad (A\cdot 3)$$

Here, $\vec{\Gamma}_L(n)$ and $\vec{\Gamma}_R(n)$ are the tunneling rates of an electron to the island through the left and right junctions, respectively. $\overleftarrow{\Gamma}_L(n)$ and $\overleftarrow{\Gamma}_R(n)$ are the tunneling rates of an electron out of the island through the left and right junctions, respectively. When there are $n$ electrons in the island, the tunneling rates can be calculated using the following equations [5]

$$\vec{\Gamma}_{L(R)}(V_{ds}, V_{gs}, n) = \frac{1}{e^2 R_{L(R)}} \frac{\vec{E}_{L(R)}(V_{ds}, V_{gs}, n)}{1 - \exp[-\vec{E}_{L(R)}(V_{ds}, V_{gs}, n)/k_B T]}, \quad (A\cdot 4.a)$$

$$\overleftarrow{\Gamma}_{L(R)}(V_{ds}, V_{gs}, n) = \vec{\Gamma}_{L(R)}(-V_{ds}, -V_{gs}, -n), \quad (A\cdot 4.b)$$

where $\vec{E}_{L(R)}(V_{ds}, V_{gs}, n)$ is the decrease in the free energy of the whole circuit due to the tunneling event through the left (right) tunnel junction, and $R_{L(R)}$ is the tunneling resistance of the left (right) tunnel junction. For the circuit in Fig. 1, $\vec{E}_{L(R)}(V_{ds}, V_{gs}, n)$ can be calculated by

$$\vec{E}_L(V_{ds}, V_{gs}, n) = \frac{e}{C_1}(Q_1 - Q_{c1}), \tag{A·5.a}$$

$$\vec{E}_R(V_{ds}, V_{gs}, n) = \frac{e}{C_2}(Q_2 - Q_{c2}), \tag{A·5.b}$$

$$Q_1 = \frac{C_1(-ne + C_2 V_{ds} + C_g V_{gs})}{C_1 + C_2 + C_g}, \tag{A·6.a}$$

$$Q_2 = \frac{C_2\left[ne + C_1 V_{ds} + C_g(V_{ds} - V_{gs})\right]}{C_1 + C_2 + C_g}, \tag{A·6.b}$$

$$Q_{c1} = \frac{C_1}{C_1 + C_2 + C_g}\frac{e}{2}, \tag{A·7.a}$$

$$Q_{c2} = \frac{C_2}{C_2 + C_1 + C_g}\frac{e}{2}, \tag{A·7.b}$$

where $Q_1$ and $Q_2$ are the charge of the left and right junctions, and $Q_{c1}$ and $Q_{c2}$ are the critical charges of the left and right junctions, respectively. At the steady state, we can calculate $p_n$ by solving $\frac{dp_n}{dt} = 0$ with the condition $\sum_{n=-\infty}^{\infty} p_n = 1$. Then, the current through the island can be calculated by

$$I = e \sum_{n=-\infty}^{\infty} p_n \left[\vec{\Gamma}_L(n) - \vec{\Gamma}_L(n)\right]. \tag{A·8}$$

For a single SEST, there is only one island. Thus, the $I_{ds}$-$V_{ds}$ and $I_{ds}$-$V_{gs}$ characteristics can be calculated by directly solving eqs. (A·1)-(A·8). However, for reconfigurable logic gates using several SESTs, it is impossible to solve the master eq. (A·1) directly because the multidimensional space of all possible charge states becomes too large. Here, we use the Monte Carlo simulation to calculate the static voltages of the output terminals as follows: the duration to the next tunneling event in a particular junction is first calculated by [4]

$$t = -\frac{\ln(r)}{\Gamma}, \tag{A·9}$$

where $r$ is an evenly distributed random number from the interval [0, 1] and $\Gamma$ is the tunneling rate given by eq. (A·4) and eqs. (A·5)-(A·7) corresponding to the considered circuit. All possible durations at any junctions are calculated, and the shortest duration is taken. After a tunneling event, island charges and island voltages also change. The calculation using eqs. (A·4)-(A·7) and (A·9) is then repeated for a few $10^5$ circles to calculate the time-averaged voltage values of the output terminals. In this way, we calculate the output voltages of the Tucker-type inverter with a variable threshold voltage, the reconfigurable AND/OR gate, and the reconfigurable logic gate for all symmetric Boolean functions at particular inputs.

**Appendix B**

Here, we calculate the TMR ratio fluctuation caused by the fluctuation of the tunnel resistances. From $R_{AP} = (TMR+1)R_P$, we obtain the relationship between the fluctuation $\Delta TMR$ of the TMR ratio and the fluctuations $\Delta R_{AP}$ and $\Delta R_P$ of tunnel resistances at AP and P magnetization configurations

$$\frac{\Delta R_{AP}}{R_{AP}} = \frac{\Delta TMR}{TMR+1} + \frac{\Delta R_P}{R_P}, \tag{B·1a}$$

or

$$\frac{\Delta TMR}{TMR} = \left(\frac{\Delta R_{AP}}{R_{AP}} - \frac{\Delta R_P}{R_P}\right)\left(1 + \frac{1}{TMR}\right). \tag{B·1b}$$

Since $\frac{\Delta R_{AP}}{R_{AP}}$ has the same sign as $\frac{\Delta R_P}{R_P}$ and $\left|\frac{\Delta R_P}{R_P}\right| \sim \left|\frac{\Delta R_{AP}}{R_{AP}}\right| \left(\equiv \left|\frac{\Delta R}{R}\right|\right)$, we have

$$\left|\frac{\Delta R_{AP}}{R_{AP}} - \frac{\Delta R_P}{R_P}\right| \sim O\left(\left|\frac{\Delta R}{R}\right|\right) \ll \frac{|\Delta R|}{R}. \tag{B·2}$$

Consequently, the maximum $|\Delta TMR|$ is given by

$$\frac{|\Delta TMR|_{max}}{TMR} = \left|\frac{\Delta R}{R}\right|\left(1 + \frac{1}{TMR}\right). \tag{B·3}$$

To satisfy $TMR - |\Delta TMR|_{max} > TMR_{min}$, we need the TMR ratio $TMR_{min}'$ to be greater than $TMR_{min}$ to overcome the resistance fluctuation:

$$TMR_{min}' = \left(TMR_{min} + \left|\frac{\Delta R}{R}\right|\right)\left(1 - \left|\frac{\Delta R}{R}\right|\right). \tag{B·4}$$

1) S. Sugahara and M. Tanaka: Appl. Phys. Lett. **84** (2004) 2307.

2) T. Matsuno, S. Sugahara and M. Tanaka: Jpn. J. Appl. Phys. **43** (2004) 6032.

3) K. K. Likharev: IEEE Trans. Magn. **MAG-23** (1987) 1142.

4) N. Kuwamura, K. Taniguchi, and C. Hamawaka: IEICE Trans. Electron. **J77-C-II** (1994) 221 [in Japanese].

5) G.-L. Ingold and Yu. V. Nazarov: in *Single Charge Tunneling* (Plenum Press, New york, 1992) Chap. 2, p. 21.

6) J. Wunderlich, T. Jungwirth, B. Kaestner, A. C. Irvine, A. B. Shick, N. Stone, K.-Y. Wang, U. Rana, A. D. Giddings, C. T. Foxon, R. P. Campion, D. A. Williams, and B. L. Gallagher: Phys. Rev. Lett. **97** (2006) 077201.

7) J. R. Tucker: J. Appl. Phys. **72** (1992) 4399.

8) H. Butler, X.-G. Zhang, T. C. Schutthess, and J. M. Maclaren: Phys. Rev. B **63** (2001) 054416.

9) X.-G. Zhang and W. H. Butler: Phys. Rev. B **70** (2004) 172407.

10) S. S. P. Parkin, C. Kaiser, A. Panchula, P. M. Rice, B. Hughes, M. Samant, and S. H. Yang: Nat. Mater. **3** (2004) 862.

11) S. Yuasa, T. Nagahama, A. Fukushima, Y. Suzuki, and K. Ando: Nat. Mater. **3** (2004) 868.

12) M. Motoyoshi, I. Yamamura, W. Ohtsuka, M. Shouji, H. Yamagishi, M. Nakamura, H. Yamada, K. Tai, T. Kikutani, T. Sagara, K. Moriyama, H. Mori, C. Fukumoto, M. Watanabe, H. Hachino, H. Kano, K. Bessho, H. Narisawa, M. Hosomi, and N. Okazaki: Symp. VLSI Technology, 2004, p. 22.

13) D. Weller, A. Moser, L. Folks, M. E. Best, W. Lee, M.F. Toney, M. Schwickert, J.-U. Thiele, and M. F. Doerner: IEEE Trans. Magn. **36** (2000) 10.

14) J. C. Slonczewski: J. Magn. Magn. Mater. **159** (1996) L1.

15) M. Hosomi, H. Yamagishi, T. Yamamoto, K. Bessho, Y. Higo, K. Yamane, H. Yamada, M. Shoji, H. Hachino, C. Fukumoto, H. Nagao, and H. Kano: IEDM Tech. Dig. 2005, p. 459.

16) K. Uchida, J. Koga, R. Ohba, and A. Toriumi: IEDM Tech. Dig. 2000, p. 863.

17) M. Saitoh, T. Murakami, and T. Hiramoto: IEEE Trans. Nanotechnol. **2** (2003) 241.

18) M. Saito and T. Hiramoto: Electron. Lett. **40** (2004) 837.
19) M. Saito and T. Hiramoto: Appl. Phys. Lett. **84** (2004) 3172.
20) M. Saitoh, H. Harata and T. Hiramoto: Jpn. J. Appl. Phys. **44** (2005) L338.
21) M. Saitoh, H. Harata and T. Hiramoto: IEDM Tech. Dig. 2004, p. 187.

**Figure Captions**

Fig. 1. (a) Schematic device structure of SEST, with FM source, drain, island, and nonmagnetic gate electrode. The magnetization directions of the electrodes are fixed and maintained parallel to each other, whereas the magnetization of the island is reversible. (b) Circuit notation of SEST.

Fig. 2. Calculated (a) $I_{ds}$-$V_{ds}$ and (b) $I_{ds}$-$V_{gs}$ characteristics in parallel (red curve) and antiparallel (blue curve) magnetization configurations of SEST. The parameters used in the simulations are $C_1$ = 9.9 aF, $C_2$ = 0.1 aF, $C_g$ = 0.1 aF, $R_{1\text{-P}}$ = 9.9 MΩ, $R_{1\text{-AP}}$ = 60.8 MΩ, $R_{2\text{-P}}$ = 0.1 MΩ, $R_{2\text{-AP}}$ = 0.95 MΩ, and $T = 0.01e^2/(C_1+C_2+C_g)k_B = 0.92$ K.

Fig. 3. Circuit configuration of Tucker-type inverter with variable threshold voltage. SEST 1 works as a p-type transistor, whereas SEST 2 works as an n-type transistor.

Fig. 4. Transfer characteristics of Tucker-type inverter with variable threshold voltage calculated by Monte Carlo simulation, when magnetization configurations of two SESTs are (SEST 1, SEST 2) = (P, AP) (blue curve) and (SEST 1, SEST 2) = (AP, P) (red curve). The parameters used in the simulations are $C_1$ = 1 aF, $C_2$ = 2 aF, $C_g$ = 8 aF, $C_b$ = 7 aF, $C_{out}$ = 24 aF, $R_{1\text{-P}} = R_{2\text{-P}}$ = 0.5 MΩ, $R_{1\text{-AP}} = R_{2\text{-AP}}$ = 5 MΩ, $V_{dd}$ = 7 mV, and $T = 0.001e^2/(C_1+C_2+C_g+C_b)k_B = 52$ mK.

Fig. 5. Circuit configuration of reconfigurable AND/OR gate. The first stage is similar to the Tucker-type inverter with a variable threshold voltage but has two digital inputs $V_A$ and $V_B$ directly coupled to each island of SESTs through two capacitors whose capacitances are half those of the Tucker-type inverter. The second stage (gray area) is a SET-based inverter working as an inverse amplifier for the output of the first stage.

Fig. 6. (a) Example of reconfigurable AND/OR gate with parameters, (b) its static AND gate operation, and (c) static OR gate operation calculated by Monte Carlo simulation. The parameters used in the simulations are $C_1 = 1$ aF, $C_2 = 2$ aF, $C_g = 8$ aF, $C_b = 7$ aF, $C_{out} = 24$ aF, $V_{dd} = 7$ mV, $T = 0.001e^2/(C_1 + C_2 + C_g + C_b)k_B = 52$ mK, $R_{1-AP} = R_{2-AP} = 5$ MΩ, and $R_{1-P} = R_{2-P} = 0.5$ MΩ for SEST 1 and SEST 2.

Fig. 7. Output voltages of (a) AND and (b) OR gate at different TMR ratios. Other parameters are the same as those in Fig. 6.

Fig. 8. Circuit configuration of reconfigurable logic gate for two-input all symmetric Boolean functions. INV 1 is a SET-based inverter with a threshold voltage higher than "1/2". INV 2 is a SET-based inverter with a threshold voltage lower than "1/2". INV 3 is a SET-based inverter with a threshold voltage of "1/2" working as an inverse amplifier for the output $V_m$ of the first stage. Two-inputs $V_A$ and $V_B$ are directly coupled to each island of SEST 1, SEST 2, INV 1, and INV 2 through two capacitors with each capacitance of $C_{in}/2$, so that the inputs $V_{in}$ to these four devices are $(V_A + V_B)/2$.

Fig. 9. (a) Example of a reconfigurable logic gate with parameters for two-input all symmetric Boolean functions, and (b) its static operation calculated by Monte Carlo simulation. The parameters used in the simulations are $C_1 = 1$ aF, $C_2 = 2$ aF, $C_g = 8$ aF, $C_b = 7$ aF, $C_{out} = 24$ aF, $V_{dd} = 7$ mV, $T = 0.001e^2/(C_1 + C_2 + C_g + C_b)k_B = 52$ mK, $R_{1-AP} = R_{2-AP} = 50$ MΩ, and $R_{1-P} = R_{2-P} = 5$ MΩ for SEST 1 and SEST 2, and $R_{3-AP} = R_{4-AP} = 500$

MΩ and $R_{3-P} = R_{4-P} = 0.5$ MΩ for SEST 3 and SEST 4.

Table I. Truth table of AND/OR circuit for each magnetization configuration.

| SEST 1 | SEST 2 | Input | | | | Output $V_{out}$ | | | Function |
|---|---|---|---|---|---|---|---|---|---|
| | | $V_A$ | 0 | 0/1 | 1 | | | | |
| | | $V_B$ | 0 | 1/0 | 1 | | | | |
| AP | P | $V_m$ | "1" | "0" | "0" | "0" | "1" | "1" | OR |
| P | AP | | "1" | "1" | "0" | "0" | "0" | "1" | AND |

Table II. Truth table of reconfigurable logic gate for all symmetric Boolean functions.

| SEST 1 | SEST 2 | SEST 3 | SEST 4 | $V_A$ $V_B$ | Input 0 0 | 0/1 1/0 | 1 1 | Output $V_{out}$ | | | Function |
|---|---|---|---|---|---|---|---|---|---|---|---|
| AP | P | AP | AP | $V_m$ | "1" | "0" | "0" | "0" | "1" | "1" | OR |
| AP | P | AP | P | | "0" | "0" | "0" | "1" | "1" | "1" | ALL 1 |
| AP | P | P | AP | | "1" | "0" | "1" | "0" | "1" | "0" | XOR |
| AP | P | P | P | | "0" | "0" | "1" | "1" | "1" | "0" | NAND |
| P | AP | AP | AP | | "1" | "1" | "0" | "0" | "0" | "1" | AND |
| P | AP | AP | P | | "0" | "1" | "0" | "1" | "0" | "1" | XNOR |
| P | AP | P | AP | | "1" | "1" | "1" | "0" | "0" | "0" | ALL 0 |
| P | AP | P | P | | "0" | "1" | "1" | "1" | "0" | "0" | NOR |

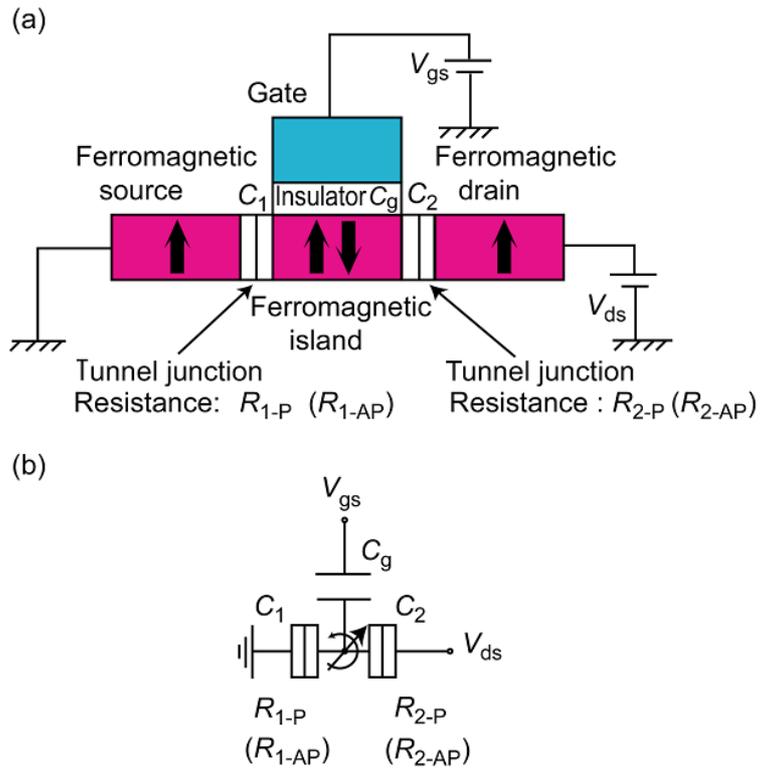

Fig. 1. Hai *et al.*

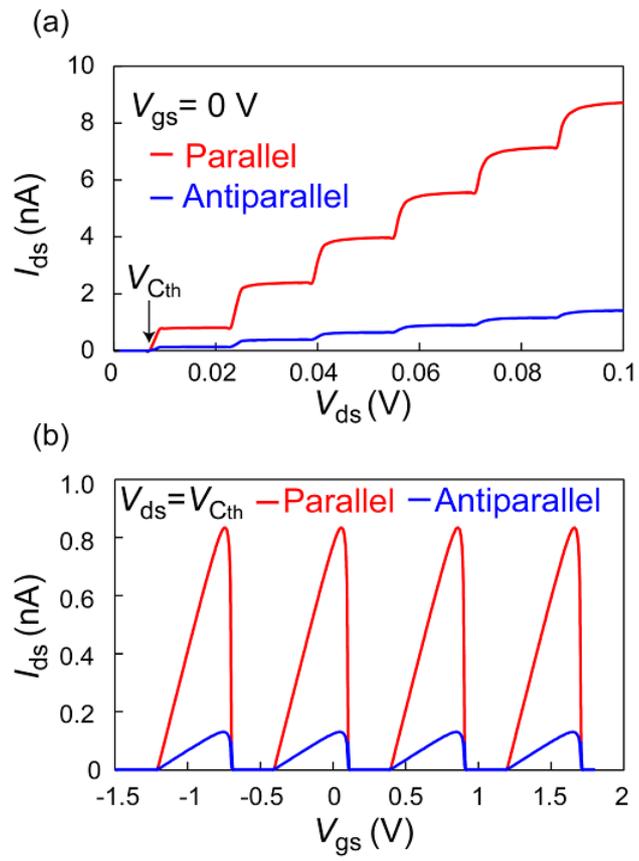

Fig. 2. Hai *et al*.

Fig. 3. Hai *et al*.

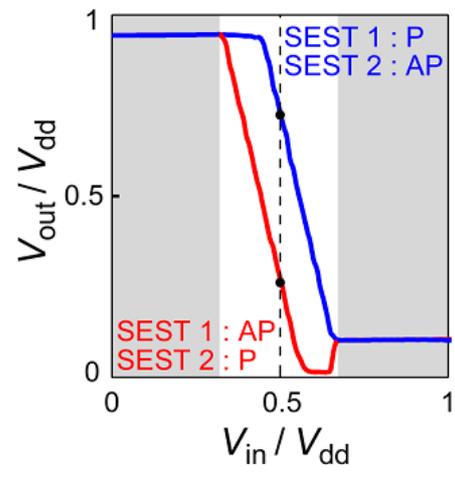

Fig. 4. Hai *et al.*

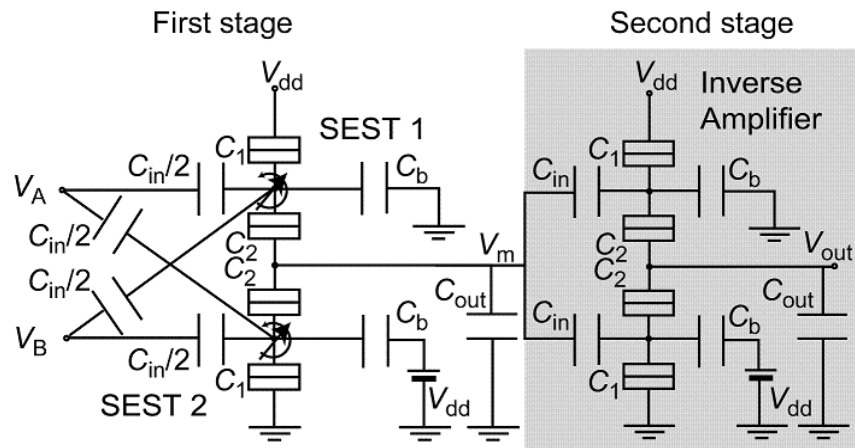

Fig. 5. Hai *et al.*

(a)

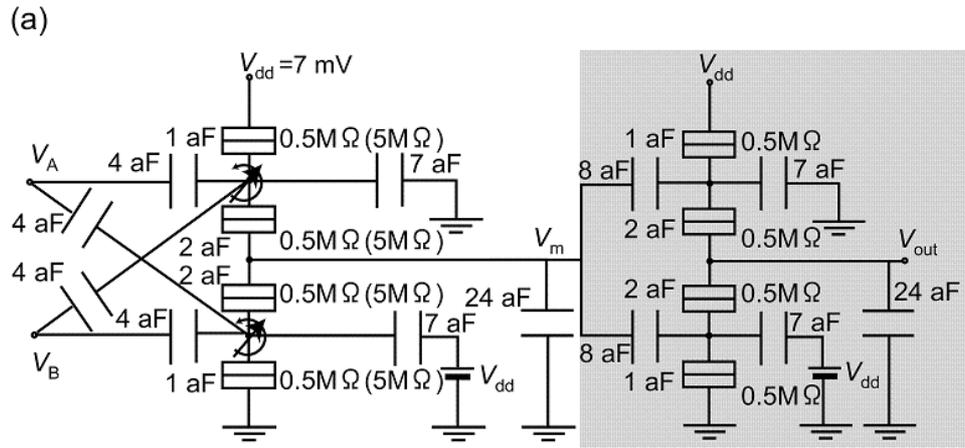

(b)

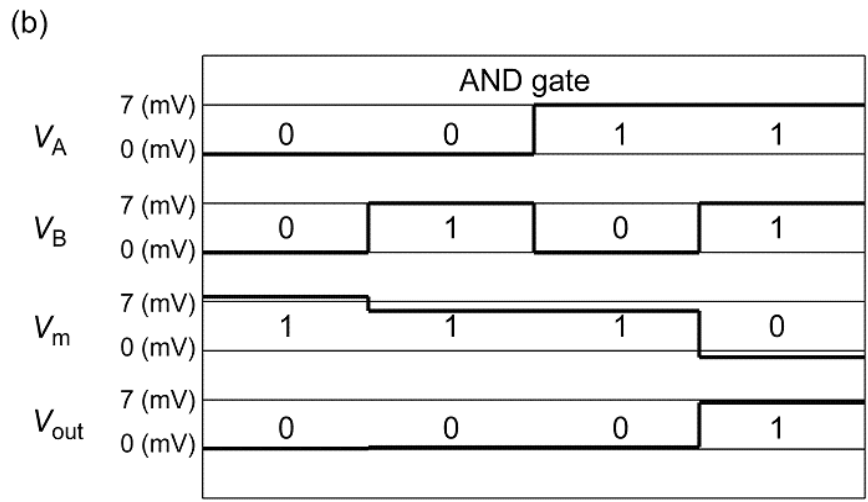

(c)

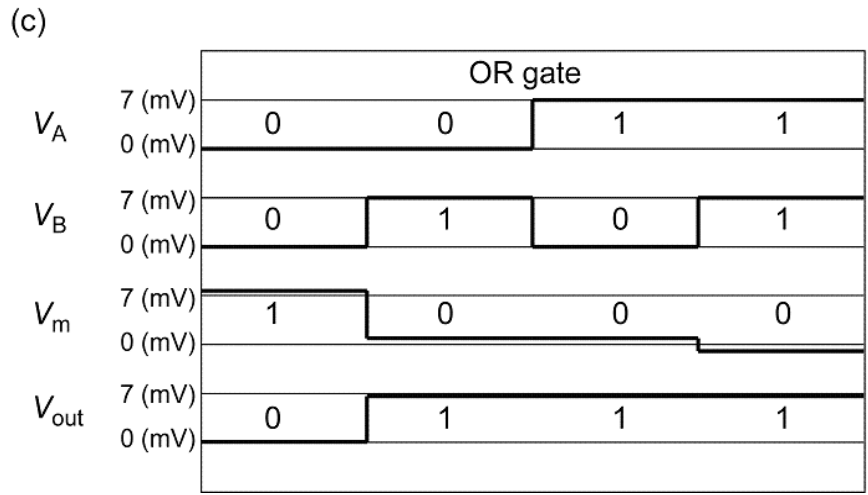

Fig. 6. Hai *et al.*

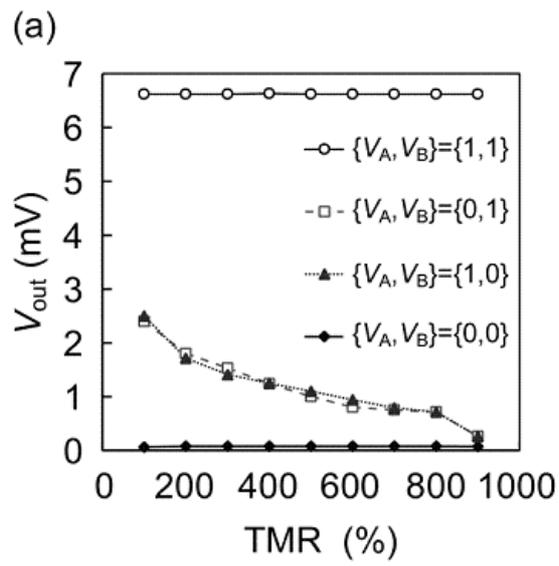

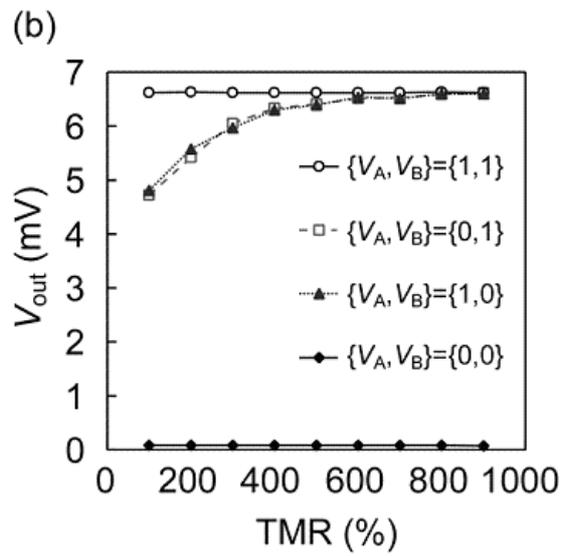

Fig. 7. Hai *et al.*

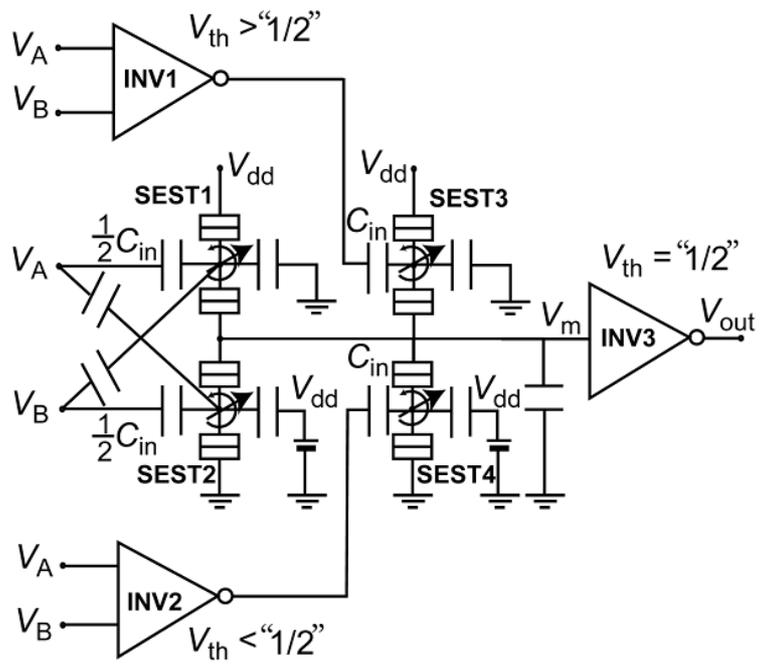

Fig. 8. Hai *et al*.

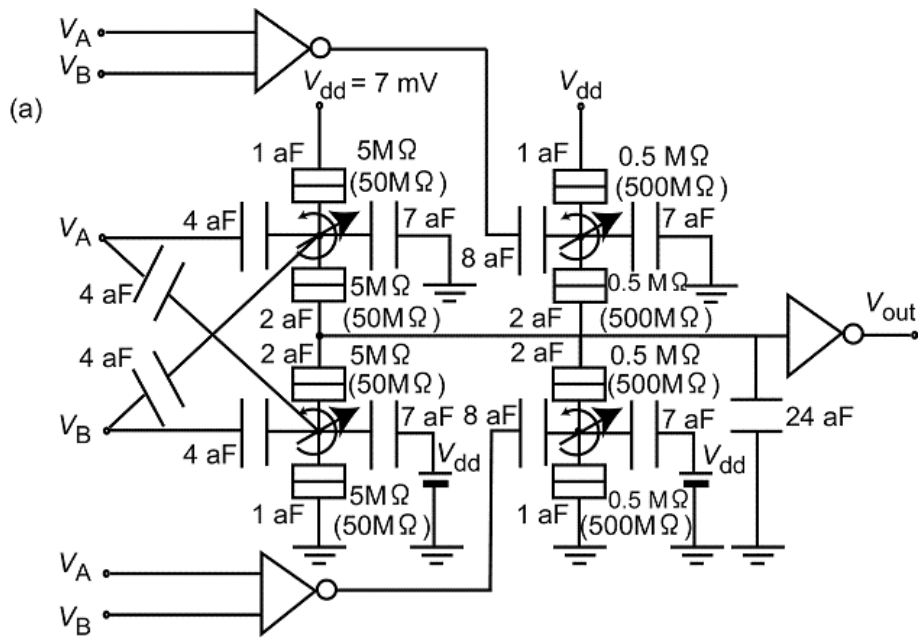

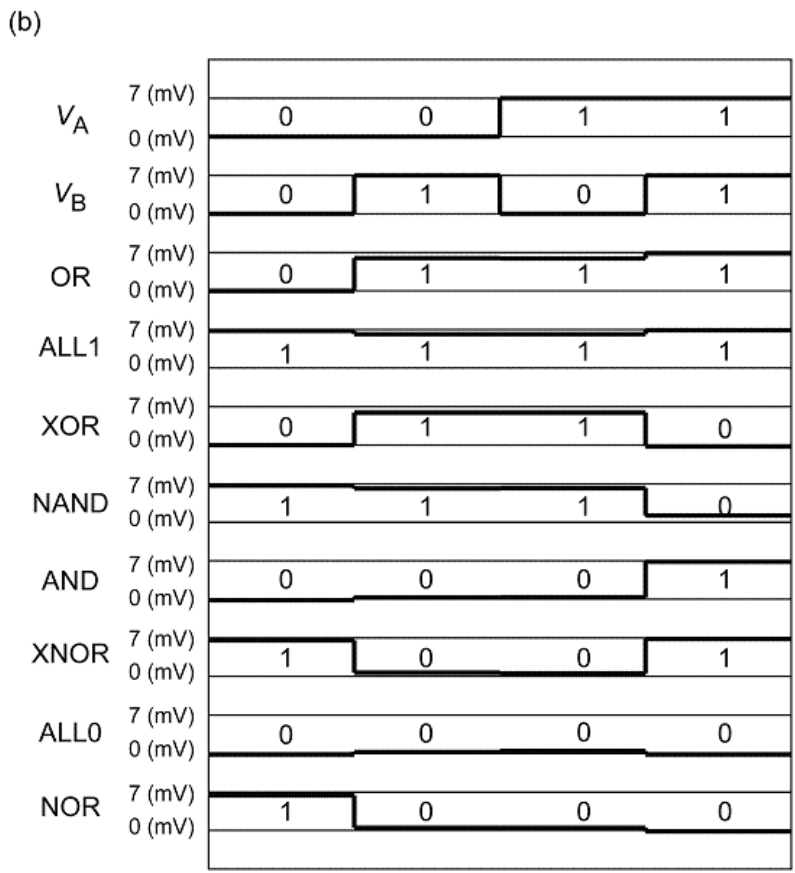

Fig. 9. Hai *et al*.